\documentclass[12pt]{article}
\usepackage{graphics,psfrag}
\usepackage{amsmath,amsthm,amssymb,epsfig,euscript,array,cite,cancel,color}
\usepackage{cases,empheq}
\usepackage[colorlinks=true]{hyperref}
\usepackage{verbatim}
\usepackage{url}
\usepackage{bbm}
   \usepackage[vcentermath]{youngtab}   
   
   \usepackage[usenames,dvipsnames]{xcolor}  

\numberwithin{equation}{section}

\theoremstyle{definition} 

\theoremstyle{definition}

\numberwithin{trial}{subsection}

\theoremstyle{remark}

\newcounter{multieqs}

\newcommand{\be}{\begin{equation}}
\newcommand{\ee}{\end{equation}}
\newcommand{\eq}[1]{(\ref{#1})}
\newcommand{\bit}{\begin{itemize}}  \newcommand{\eit}{\end{itemize}}

\newcommand{\bm}[1]{\mbox{\boldmath $#1$}}
\newcommand{\rf}[1]{(\ref{#1})}

\def\bd{\begin{document}}
\def\ed{\end{document}}
\def\nn{\nonumber}
\def\bea{\begin{eqnarray}}
\def\eea{\end{eqnarray}}
\let\bm=\bibitem

\def\la{\langle}
\def\ra{\rangle}

\def\npb#1#2#3{Nucl. Phys. {\bf{B#1}} #3 (#2)}
\def\plb#1#2#3{Phys. Lett. {\bf{#1B}} #3 (#2)}
\def\prl#1#2#3{Phys. Rev. Lett. {\bf{#1}} #3 (#2)}
\def\prd#1#2#3{Phys. Rev. {D \bf{#1}} #3 (#2)}
\def\cmp#1#2#3{Comm. Math. Phys. {\bf{#1}} #3 (#2)}
\def\cqg#1#2#3{Class. Quantum Grav. {\bf{#1}} #3 (#2)}
\def\nppsa#1#2#3{Nucl. Phys. B (Proc. Suppl.) {\bf{#1A}}#3 (#2)}
\def\ap#1#2#3{Ann. of Phys. {\bf{#1}} #3 (#2)}
\def\ijmp#1#2#3{Int. J. Mod. Phys. {\bf{A#1}} #3 (#2)}
\def\rmp#1#2#3{Rev. Mod. Phys. {\bf{#1}} #3 (#2)}
\def\mpla#1#2#3{Mod. Phys. Lett. {\bf A#1} #3 (#2)}
\def\jhep#1#2#3{J. High Energy Phys. {\bf #1} #3 (#2)}
\def\atmp#1#2#3{Adv. Theor. Math. Phys. {\bf #1} #3 (#2)}

\def\N{{\cal N}}
\def\sst{\scriptscriptstyle}
\def\thetabar{\bar\theta}
\def\Tr{{\rm Tr}}
\def\one{\mbox{1 \kern-.59em {\rm l}}}

\def\a{\alpha}      \def\da{{\dot\alpha}}  \def\dA{{\dot A}}
\def\b{\beta}       \def\db{{\dot\beta}}  
\def\g{\gamma}  \def\G{\Gamma}  \def\dc{{\dot\gamma}}  
\def\d{\delta}  \def\D{\Delta}  \def\ddt{\dot\delta}  
\def\e{\epsilon}        \def\ve{\varepsilon}  
\def\f{\phi}    \def\F{\Phi}    \def\vvf{\f}  
\def\h{\eta}  
\def\k{\kappa}  
\def\l{{\lambda}} \def\L{\Lambda}  
\def\m{\mu} \def\n{\nu}  
\def\o{\omega}  
\def\p{\pi} \def\P{\Pi}  
\def\r{\rho}  
\def\s{\sigma}  \def\S{\Sigma}  
\def\t{\tau}  
\def\th{\theta} \def\Th{\Theta} \def\vth{\vartheta}  
\def\X{\Xeta}  
\def\z{\zeta}  

\def\na{\nabla}  

\def\cA{{\cal A}} \def\cB{{\cal B}} \def\cC{{\cal C}}  
\def\cD{{\cal D}} \def\cE{{\cal E}} \def\cF{{\cal F}}  
\def\cG{{\cal G}} \def\cH{{\cal H}} \def\cI{{\cal I}}  
\def\cJ{{\cal J}} \def\cK{{\cal K}} \def\cL{{\cal L}}  
\def\cM{{\cal M}} \def\cN{{\cal N}} \def\cO{{\cal O}}  
\def\cP{{\cal P}} \def\cQ{{\cal Q}} \def\cR{{\cal R}}  
\def\cS{{\cal S}} \def\cT{{\cal T}} \def\cU{{\cal U}}  
\def\cV{{\cal V}} \def\cW{{\cal W}} \def\cX{{\cal X}}  
\def\cY{{\cal Y}} \def\cZ{{\cal Z}}

\def\ua{{\underline{\alpha}}} 
 \def\ub{\underline{\phantom{\alpha}}\!\!\!\beta}  
\def\uc{\underline{\phantom{\alpha}}\!\!\!\gamma}  
\def\um {{\underline{\mu}}}  
\def\ud{{\underline{\delta}}} 
\def\ue{\underline\epsilon}  
\def\una{{\underline a}}\def\uA{{\underline A}}  
\def\unb{{\underline b}}\def\uB{{\underline B}} 
\def\unc{{\underline c}}\def\uC{{\underline C}}  
\def\und{{\underline d}}\def\uD{{\underline D}}  
\def\une{{\underline e}}\def\uE{{\underline E}}  
\def\unf{{\underline{\phantom{e}}\!\!\!\! f}}\def\uF{{\underline F}}  
\def\unm{{\underline m}\def\uM{\underline M}} 
\def\unn{{\underline n}\def\uN{\underline N}} 
\def\unp{{\underline{\phantom{a}}\!\!\! p}}\def\uP{{\underline P}}  
\def\unq{{\underline{\phantom{a}}\!\!\! q}}  
\def\uQ{{\underline{\phantom{A}}\!\!\!\! Q}}  
\def\uH{{\underline{H}}}  
\def\uM{{\underline{M}}}
\def\uN{{\underline{N}}}
\def\unl{{\underline{l}}}

\def\Ah{{\hat{A}}}  
\def\Dh{{\hat{D}}}
\def\Gh{{\hat{G}}}
\def\Fh{{\hat{F}}}
\def\Ih{{\hat{I}}} 
\def\Jh{{\hat{J}}} 
\def\Kh{{\hat{K}}}
\def\Lh{{\hat{L}}} 
\def\Ph{{\hat{P}}}
\def\Rh{{\hat{R}}}
\def\Vh{{\hat{V}}} 
\def\Xh{{\hat{X}}}
 
\def\ah{{\hat{a}}}
\def\bh{{\hat{b}}}
\def\ch{{\hat{c}}}
\def\gh{{\hat{g}}}
\def\dh{{\hat{d}}}
\def\hh{{\hat{h}}}
\def\uh{{\hat{u}}}  
\def\vh{{\hat{v}}}
\def\xh{{\hat{x}}} 
\def\yh{{\hat{y}}}
\def\zh{{\hat{z}}}
\def\ph{{\hat{p}}}
\def\qh{{\hat{q}}}
\def\thh{{\hat{t}}}  
\def\xih{\hat{\xi}}  
\def\Psih{\hat{\Psi}}    
\def\mh{{\hat{m}}}
\def\nh{{\hat{n}}}
\def\ih{{\hat{i}}}
\def\jh{{\hat{j}}}
\def\kh{{\hat{k}}}
\def\aah{{\hat{\alpha}}}
\def\bbh{{\hat{\beta}}}
\def\ggh{{\hat{\gamma}}}
\def\llh{{\hat{\ell}}} 
\def\ph{{\hat{p}}}

\def\psit{\tilde{\psi}}  
\def\Psit{\tilde{\Psi}}   
\def\Psibt{\tilde{\bar{Psi}}}  

\def\st{\tilde{\sigma}}  

\def\delt{\tilde{\delta}}
\def\Phit{\tilde{\Phi}}   
\def\Phitb{\overline{\tilde{Phi}}}  
\def\tht{\tilde{\th}}  
\def\lt{\tilde{\l}}
\def\chit{\tilde{\chi}}   
\def\phit{\tilde{\phi}} 

\def\At{\tilde{A}}
\def\Bt{\tilde{B}}
\def\Ct{\tilde{C}}
\def\Dt{\tilde{D}}
\def\Et{\tilde{E}}
\def\Ft{\tilde{F}}
\def\Gt{\tilde{G}}
\def\Ht{\tilde{H}}
\def\It{\tilde{I}}
\def\Jt{\tilde{J}}
\def\Qt{\tilde{Q}}  
\def\Rt{\tilde{R}}  
\def\Mt{\tilde{M }}  
\def\Nt{\tilde{N}}   
\def\St{\tilde{S}}
\def\Vt{\tilde{V}}
\def\Xt{\tilde{X}} 
\def\at{\tilde{a}}
\def\ct{\tilde{c}}
\def\dt{\tilde{d}}
\def\htt{\tilde{h}} 
\def\ft{\tilde{f}}
\def\gt{\tilde{g}}
\def\pt{\tilde{p}}  
\def\qt{\tilde{q}}  
\def\vt{\tilde{v}}  
\def\nt{\tilde{n}}  
\def\ut{\tilde{u}}  
\def\wt{\tilde{w}}  
\def\zt{\tilde{z}} 
\def\xt{\tilde{x}} 
\def\yt{\tilde{y}} 
\def\Psit{\tilde{\Psi}}
\def\vphit{\tilde{\varphi}}

\def\gamt{\tilde{\gamma}}

\def\Tt{\tilde{T}}

\def\eb{\bar{\epsilon}} 
\def\delb{\bar{\partial}}  
\def\thb{\bar{\theta}}
\def\Thb{{\bar{\Theta}}}
\def\mub{\bar{\mu}}
\def\lamb{\bar{\l}}
\def\psib{\bar{\psi}}
\def\sb{\bar{\sigma}}
\def\xib{\bar{\xi}}
\def\chib{\bar{\chi}}

\def\Psib{\bar{\Psi}}
\def\Phib{\bar{\Phi}}
\def\Lamb{\bar{\Lambda}}
\def\Sb{{\overline \Sigma}}
\def\cb{\bar{c}}
\def\hb{\bar{h}}
\def\qb{\bar{q}}
\def\wb{\bar{w}}
\def\zb{{\bar{z}}}
\def\Hb{\bar{H}}
\def\Qb{{\bar Q}}
\def\Omegab{\overline{\Omega}}
\def\ob{\overline{\omega}}
\def\Gab{{\bar{\Gamma}}}

\def\Ab{{\overline A}} \def\Bb{{\overline B}} \def\Cb{{\overline C}}  
\def\Db{{\overline D}} \def\Eb{{\overline E}} \def\Fb{{\overline F}}  
\def\Gb{{\overline G}} 
\def\Ib{{\overline I}}  
\def\Jb{{\overline J}} \def\Kb{{\overline K}} \def\Lb{{\overline L}}  
\def\Mb{{\overline M}} \def\Nb{{\overline N}} \def\Ob{{\overline O}}  
\def\Pb{{\overline P}}  \def\Rb{{\overline R}}  
 \def\Tb{{\overline T}} \def\Ub{{\overline U}}  
\def\Vb{{\overline V}} \def\Wb{{\overline W}} \def\Xb{{\overline X}}  
\def\Yb{{\overline Y}} \def\Zb{{\overline Z}}  

\def\fb{{\overline f}}
\def\gb{{\overline g}}
\def\mb{{\overline m}}
\def\lb{{\overline l}}
\def\yb{{\overline y}}
  
\def\ldel{{\overleftarrow{\del}}}
\def\rdel{{\overrightarrow{\del}}}
\def\ldeldel{{\overleftarrow{\del^2}}}
\def\rdeldel{{\overrightarrow{\del^2}}}
\def\ldelb{{\overleftarrow{\bar{\del}}}}
\def\rdelb{{\overrightarrow{\bar{\del}}}}

\def\bone{{\bf 1}}  

\def\va{{\vec a}}
\def\vk{{\vec k}}
\def\vp{{\vec p}}
\def\vq{{\vec q}}
\def\vx{{\vec x}}
\def\vy{{\vec y}}
\def\vu{{\vec u}}
\def\vv{{\vec v}}

\def\vs{{\vec \sigma}}
\def\vtau{{\vec \tau}}

\newcommand{\ov}[1]{\overrightarrow{#1}}

\def\d{\delta}\def\D{\Delta}\def\ddt{\dot\delta}  
  
\def\pa{\partial} \def\del{\partial}  
\def\xx{\times}  
\def\uno{\mbox{1 \kern-.59em {\rm l}}}    
  
\def\trp{^{\top}}  
\def\inv{^{-1}}  
\def\dag{{^{\dagger}}}  
\def\pr{^{\prime}}  
  
\def\rar{\rightarrow}  
\def\lar{\leftarrow}  
\def\lrar{\leftrightarrow}  
  
\newcommand{\0}{\,\!}      
\def\one{1\!\!1\,\,}  
\def\im{\imath}  
\def\jm{\jmath}  
  
\newcommand{\tr}{\mbox{tr}}  
\newcommand{\slsh}[1]{/ \!\!\!\! #1}  
  
\def\vac{|0\rangle}  
\def\lvac{\langle 0|}  
  
\def\hlf{\frac{1}{2}}  
\def\ove#1{\frac{1}{#1}}  

\def\Box{\square}  
\def\CC {\mathbb{C}}
\def\FF {\mathbb{F}}
\def\RR{\mathbb{R}}
\def\NN{\mathbb{N}}  
\def\ZZ{\mathbb{Z}}  
\def\bb#1{{\bf #1}}  
\def\bcomment#1{}  
\def\bfhat#1{{\bf \hat{#1}}}  
\def\VEV#1{\left\langle #1\right\rangle}  

\newcommand{\ex}[1]{{\rm e}^{#1}} \def\ii{{\rm i}}  

\newcommand{\lrbrk}[1]{\left(#1\right)}
\newcommand{\lrsbrk}[1]{\left[#1\right]}
\newcommand{\lrcbrk}[1]{\left\{#1\right\}}
\newcommand{\sfrac}[2]{{\textstyle\frac{#1}{#2}}}
 
\def\stw{{\sqrt{2}}}

\def\rf {{\rm f}}
\def\ri {{\rm i}}
\def\rj {{\rm j}}
\def\rk {{\rm k}}
\def\rl {{\rm l}}
\def\rs {{\scriptscriptstyle \rm S}}
\def\rt {{\scriptscriptstyle \rm T}}

\def\rQ {{\scriptscriptstyle \rm \cQ}}
\def\rR {{\scriptscriptstyle \rm \cR}}

\def\cQb{{\cal \Qb}}
\def\cRb{{\cal \Rb}}
\def\cWb{{\cal \Wb}}

\def\fd {{\rm N}}
\def\afd {{\overline{\rm N}}}

\def \II {I\hspace{-.1em}I\hspace{.1em}}
\def \IIA {\mbox{\II A\hspace{.2em}}}
\def \IIB {\mbox{\II B\hspace{.2em}}}
\def \gs {g^s}
\def \ls {\lambda^s}

\def \I {{\cal I}}
\def \qs {q\hspace{-.53em}/\hspace{.15em}}
\def \ks {k\hspace{-.53em}/\hspace{.15em}}
\def \YM {{\mbox{\tiny YM}}}
\def \gym {g_{\YM}}

\def \Lc {\L_c}
\def\IR{\relax{\rm I\kern-.18em R}}
\def \id {{\bf 1}}

\def\cci{\ell}
\def\ccj{\ell'}

\def \thbb{\overline{\th\th}}
\newcommand \ol{\overline}
\def \lamb{\bar{\lambda}}
\def \vphi{\varphi}
\def \lambh{\hat{\bar{\lambda}}}
\def \lh{\hat{\lambda}}
\def \dd{\ddagger}

\def \ad {\dot{a}}
\def \bd {\dot{b}}
\def \cd {\dot{c}}
\def  \ddd {\dot{d}}
\def \ed {\dot{e}}
\def \fd {\dot{f}}
\def \Bh {\hat{B}}
\def \zm {{(0)}}
\def \nz {{(\text{KK})}}
\def \3{{(3)}}
\def \diag {\text{diag}}
\def \inm {{(m^{-1})}}
\def \3{{(3)}} 
\def \6{{(6)}}
\def \2{{(2)}}
\def \7{{(7)}} 
\def \4{{(4)}}
\def\1{{(1)}}
\def\5{{(5)}}
\def\0{{(0)}}

\def \DBI{{\text{DBI}}}
\def\et{{\tilde{\e}}}
\def\w{{\wedge}}
\def\bbV{{\mathbb{V}}}
\def\M{{(\text{M})}}
\def\T{{(\text{T})}}

\def\Hbt{{\tilde{\bar{H}}}}
\def\Fbt{{\tilde{\bar{F}}}}

\colorlet{1}{red}
\colorlet{2}{green}
\colorlet{3}{blue}
\colorlet{4}{cyan}

\def\PI{{P^\perp}}
\def\Pt{{\tilde{P}}}
\def\PIt{{\tilde{P}^\perp}}
\def\cPI{{\cP^\perp}}
\def\cPt{{\tilde{\cP}}}
\def\cPIt{{\tilde{\cP}^\perp}}
\def\bigw{{\bigwedge}}

\author{Sheng-Lan Ko\footnote{ko.shenglan@gmail.com}~
and  Pichet Vanichchapongjaroen\footnote{pichetv@nu.ac.th}~
\\
\\
{\small  
	\it The Institute for Fundamental Study ``The Tah Poe Academia Institute",}
\\
{\small\it Naresuan University, Phitsanulok 65000, Thailand}
}

\title{\bf A covariantisation of M5-brane action in dual formulation}

\begin{document}
\maketitle

\abstract{
We construct a manifestly diffeomorphic M5-brane action
in dual formulation coupled to an eleven-dimensional supergravity
target space. The covariantisation is carried out by using (generalised) PST technique with 
$5$ auxiliary scalar fields,
which
are obtained by using a geometrical consideration
as a
reduction of an auxiliary $4-$form of
Maznytsia-Preitschopf-Sorokin.
As is typical in PST-covariantised theory,
our construction possesses as usual the two local PST symmetries.
By using one of the local PST symmetries, the action can be reduced
to the non-manifestly covariant M5-brane action
in dual formulation constructed earlier by the authors.
The discussion on double dimensional reduction to D4-brane,
and on the comparison of on-shell action then easily follows.
}

\thispagestyle{empty}
\newpage
\tableofcontents

\section{Introduction}
On a quest to construct an M5-brane action in the literature,  
one of the important obstacles is related to
a certain field content of an M5-brane ---
a chiral 2-form field, i.e.
a 2-form gauge field with non-linear self-dual 3-form field strength.
The obstacle is evident on the theory of a chiral 2-form field
by itself,
or more generally on
the theory of a chiral $p-$form in $(2p+2)$ dimensions
for even $p$.
On these theories,
it is non-trivial to impose
the Lorentz invariance
and the self-duality together at the action level.
As a related issue, at the linear level,
self-duality conditions are first order differential equations,
which are not obtainable via a standard consideration
from a quadratic action.
There are two ways to resolve these issues.
For definiteness, let us focus the discussion on chiral 2-form theories.

The first way is to
give up the manifest $SO(1,5)$
Lorentz invariance at the action level
by making a split of spacetime.
This way was achieved by
\cite{Henneaux:1988gg},
which presented the chiral 2-form action
with manifest $SO(5)$ subgroup of $SO(1,5)$
Lorentz symmetry,
and alternatively by \cite{Perry:1996mk},
in which the action possesses
a manifest $SO(1,4)$ subgroup of
$SO(1,5)$ Lorentz symmetry.

Alternatively,
as was done by \cite{Pasti:1995ii, Pasti:1996va,Pasti:1996vs},
a manifestly Lorentz covariant
chiral 2-form action can be constructed
by introducing an auxiliary scalar field,
which appears non-linearly even in the quadratic theory.
This way of introducing the auxiliary scalar field
is called the PST covariantisation,
and the theory itself is known as the PST theory.
This theory possesses two notable local symmetries,
one of which is used to ensure the field equation
reduces to self-duality condition,
whereas the other is used to
ensure the auxiliary nature of the auxiliary field.
The latter symmetry is used to gauge fix
the auxiliary field, reducing the
theory to the non-manifest covariant versions \cite{Henneaux:1988gg, Perry:1996mk}, 
thus realising non-covariant theories as different gauge-fixings of the PST theory.

From either way,
the chiral 2-form theory
can be extended
to the complete M5-brane theory
coupled to 11d supergravity background \cite{Pasti:1997gx,Bandos:1997ui,Aganagic:1997zq}.
This makes use of Green-Schwarz formalism \cite{Green:1983wt},
in which an M5-brane is coupled
to the supersymmetric background.
This is shown by \cite{Howe:1996yn,Bandos:1997gm, Howe:1997vn}
that the field equations agree with those obtained
from the superembedding approach \cite{Sorokin:1999jx},
in which a supersymmetric M5-brane is coupled
to the supersymmetric background.

In the framework of string theory and M-theory,
theories have been known and expected to be related
to one another by some duality transformations.
In the case of M5-brane theory,
an early attempt of the dualisation
is given by \cite{Maznytsia:1998xw,Maznytsia:1998yr}
in which a quadratic PST 
covariantised chiral 2-form action
is dualised. It was found that
while the dualisation applied to the chiral 2-form
does not change the theory,
the dualisation applied to the
auxiliary field gives rise to a
quadratic PST covariantised chiral 2-form action
with an auxiliary 4-form.
This theory is said to be in a dual formulation.

The paper \cite{Ko:2016cpw}
extended this theory
to a complete M5-brane theory coupled to the 11d supergravity background.
The construction did not make use of the auxiliary field,
but instead made the split in the worldvolume indices
in such a way that
the action only presented a manifest 5d worldvolume diffeomorphism,
but it can be shown that there is an off-shell modified
6d worldvolume diffeomorphism.

In the standard formulation, the non-manifestly covariant
M5-brane theory contains second-class constraints  
which complicate the quantisation.
A way to remedy this is to make the PST covariantisation,
giving PST-covariant M5-brane theory
containing only first-class constraints \cite{Pasti:1996vs}.
We expect this to be analogous to the dual formulation.
Although the complete M5-brane theory in dual formulation
has been constructed, the theory is still not manifestly
covariant. The covariantisation of this theory is then
expected to put the action in the form which makes
it simpler to later carry out the quantisation procedure.
In fact, the covariantisation of 6d chiral 2-form theory
with quadratic action was already given in \cite{Maznytsia:1998xw,Maznytsia:1998yr}
by using an auxiliary 4-form.
However, by a closer inspection
it turns out that there seems to be a potential
issue which might prevent the extension
to the complete M5-brane theory.

The goal of this paper is to construct a covariant
complete M5-brane theory in dual formulation.
As to be discussed
in this paper, it is
still inconclusive whether
using an auxiliary 4-form would really lead to an issue.
Instead of keeping on investigating to see whether the issue
truly exists, we simply aim to
look for a special case of auxiliary field which make it possible to construct a covariant
complete M5-brane theory in dual formulation.
It turns out 
that the covariantisation and extension to the complete M5-brane theory
can be made possible by using 5 auxiliary scalar fields.

On the technical side, the constructions
and studies in this paper
are made possible using differential form language.
In particular, the 5 auxiliary scalar fields
appear in the theory via projector matrices,
which can be incorporated into the differential form language
through the use of the induced linear transformation,
to be given a quick review in this paper.

This paper is organised as follows.
Section \ref{sec:M5Ss} starts by reviewing the standard M5-brane theory,
then followed by the main result of this paper,
which is M5-brane theory in dual formulation.
Section \ref{sec:derive} presents the derivation,
by first reviewing 6d chiral 2-form theory
with quadratic action covariantised using an auxiliary $4-$form \cite{Maznytsia:1998xw,Maznytsia:1998yr},
and stating its potential issues.
Then motivates an alternative way to covariantise,
which is by the use of $5$ auxiliary scalars.
Finally we proceed to make a detailed analysis in quadratic action,
and nonlinear action.
Section \ref{sec:gaugefixtononcov} discusses that
the covariantised M5-brane action in dual formulation
using $5$ auxiliary scalar fields
is reduced, upon a suitable gauge-fixing of the auxiliary fields,
to the action constructed in \cite{Ko:2016cpw}.
Finally, in section \ref{sec:conclude} we give conclusions
and suggestions for future works.

\section{The M5-brane actions}  \label{sec:M5Ss}
In this paper, the 
11--dimensional target superspace is
parametrised by supercoordinates $Z^{\mathcal M}=(X^M,\th)$, 
in which $X^M$ are eleven bosonic coordinates 
and $\th$ are 32 real fermionic coordinates. 
The geometry of the 11d supergravity are described by tangent-space vector super-vielbeins $E^A(Z)=dZ^{\cM}E_{\cM}{}^A(Z)$ ($A=0,1,2,\cdots, 10)$ and Majorana-spinor super-vielbeins $E^\alpha(Z)=dZ^{\cM}E_{\cM}{}^\alpha(Z)$ ($\a=1,2,\cdots, 32)$.
The kappa-symmetry of the $M5$-brane action
requires that
the vector super-vielbein satisfies the torsion constraint
\be\label{T}
T^A=DE^A=dE^A+E^B\Omega_B{}^A=-iE^\a \Gamma^A_{\a\b}E^\beta\,,
\ee
where $\Omega_B{}^A(Z)$ is the 1-form spin connection in eleven dimension, $\Gamma^A_{\a\b}=\Gamma^A_{\b\a}$ are real symmetric gamma matrices and the exterior differential acts from the right.
The signature of the metric is taken to be mostly plus.

The M5-brane worldvolume is parametrised by the coordinates $x^\mu$ ($\mu=0,1,\cdots, 5$).
Its induced metric is constructed with the pull-backs of the vector super-vielbeins $E^A(Z)$
\be\label{g6}
g_{\mu\nu}(x)=E_\m^AE^B_\n
\eta_{AB},
\qquad E_\mu^A=\partial_\m Z^{\cN}E_{\cN}{}^A(Z(x)).
\ee
It couples to the 11d supergravity 3-form gauge superfield, $C_3(Z)=\frac 1{3!}
dZ^{\cM_1}dZ^{\mathcal M_2}dZ^{\mathcal M_3}
C_{\cM_3  \cM_2  \cM_1}$,
and its $C_6(Z)$ dual.
Their field strengths are constrained as follows
\be \label{F47}
\begin{split}
dC_3
&= -\frac i2 E^AE^BE^\alpha E^\beta(\Gamma_{BA})_{\alpha\beta}+\frac 1{4!} E^AE^BE^CE^DF^{(4)}_{DCBA}(Z)\,, \\
dC_6-C_3dC_3
&= \frac{2i}{5!} E^{A_1}\cdots E^{A_5}E^\alpha E^\beta(\Gamma_{A_5\cdots A_1})_{\alpha\beta}+\frac 1{7!} E^{A_1}\cdots E^{A_7}F^{(7)}_{A_7\cdots A_1}(Z)\,\\
F^{(7)\,A_1\cdots A_7}
&= \frac 1{4!}\epsilon^{A_1\cdots A_{11}}F^{(4)}_{A_8\cdots A_{11}}\,, \qquad \epsilon^{0...10}=-\epsilon_{0...10}=1. 
\end{split}
\ee
The M5-brane carries the chiral 2-form gauge field 
$B_2(x)=\ove{2}dx^\mu dx^\nu B_{\n\m}(x)$
with field strength
\be\label{H3}
H_3=dB_2+C_3\,,
\ee
where $C_3(Z(x))$ is the pullback of the 3-form gauge field on the M5-brane worldvolume.

\subsection{PST-covariantised M5--brane action}\label{subsecPSTM5}
The original M5-brane action in a generic $D=11$ supergravity superbackground
is constructed in \cite{Pasti:1997gx,Bandos:1997ui,Aganagic:1997zq}.
In order for the worldvolume theory to be manifestly covariant
at the action level, an auxiliary scalar field
$a(x)$ is introduced.
Its gradient $\pa_\m a$ could be either time-like or space-like.
These
two cases share the same action.
However, for definiteness, we present the action
in the form which accommodates space-like case:
\bea \label{PSTM5}
S_{PST-M5}
&=& -\int_{\mathcal{M}_6} d^6x \lrsbrk{\sqrt{ - \det\lrbrk{g_{\m\n} + i (\Ht\cdot u)_{\m\n}}}  + \frac{\sqrt{-g}}{4} (\Ht\cdot u)^{\m\n}(H\cdot u)_{\m\n}} \nn\\
&&
 + \ove{2}\int_{\mathcal{M}_6} \lrbrk{ C_6 + H_3 \wedge C_3 },
\eea
where
\be
(H\cdot u)_{\m\n} = H_{\m\n\r}u^\r,\qquad
(\Ht\cdot u)_{\m\n} = \Ht_{\m\n\r}u^\r,\qquad
u_\r = \frac{\pa_\r a}{\sqrt{\pa_\m a g^{\m\n}\pa_\n a}},
\ee
\be   \label{oldHbt}
\Ht^{\r\m\n} \equiv \frac{1}{6\sqrt{-g}} \, \e^{\r\m\n\l\s\t} H_{\l\s\t},\quad
g=\det g_{\m\n}\,,
\ee
with
$$
\epsilon^{0\cdots 5}=-\epsilon_{0\cdots 5}=1\,.
$$

In addition to the conventional abelian gauge symmetry for the chiral 2-form,
the action (\ref{PSTM5}) has also the following two local gauge symmetries.
The first one, of type called PST1, is given by
\be \label{15SYM1}
\d B_{\mu\nu} = 2\del_{[\mu} a\, \Phi_{\nu]}(x), \qquad \d a(x) = 0,
\ee
with $\Phi_\mu(x)$ being arbitrary local functions on the woldvolume.
This symmetry ensures that the equation of motion of $B_2$ reduces to the non--linear self--duality condition
\be\label{sdM5o}
(H\cdot u)_{\mu\nu}= \cU_{\mu\nu}
\,,
\ee
where
\be\label{calV}
\cU^{\mu\nu}
\equiv -2\,\frac{\d \sqrt{\det(\delta^\nu_{\mu} + i(\Ht\cdot u)_{\mu}{}^{\nu})}}{\d (\Ht\cdot u)_{\mu\nu}}.
\ee
Another local gauge symmetry, whose type is called PST2, is given by
\be \label{15SYM2}
\d a = \vphi(x),\qquad \d B_{\mu\nu} = \frac{\vphi(x)}{\sqrt{(\del a)^2}} ( H_{\mu\nu} - \cU_{\mu\nu}),
\ee
with $\vphi(x)$ being an arbitrary local function on the woldvolume.
This symmetry ensures that the scalar field $a(x)$ is indeed arbitrary
and that the action is $6d$ covariant.

The action \eqref{PSTM5} is also invariant under the local fermionic kappa--symmetry transformations
which acts on the worldvolume fields and pullbacks of the target-space fields
as follows
\bea\label{PSTkappa}
i_\k E^\alpha \equiv \delta_\k Z^{\mathcal M} E^\alpha_{\mathcal M} = \frac 12 (1+\bar\Gamma)^\alpha{}_{\beta} \k^{\beta}, \quad
i_\k E^A \equiv \delta_\k Z^{\mathcal M} E^A_{\mathcal M}  =  0, \\
\delta_\k g_{\mu\nu} = -4i E^\alpha_{(\mu} (\G_{\nu)})_{\alpha\beta} \, i_\k E^{\beta}, \quad
\d_\k{H}^\3 = i_\k d C^\3, \quad \delta_\k a(x) = 0\,,    \nn
\eea
where $\k(x)$ is the parameter of kappa--symmetry transformation.
The matrix $\bar\G$ is given by
\be\label{barGPST}
\begin{split}
\sqrt{\det(\delta_{\mu}^{\nu} + i(\Ht\cdot u)_{\mu}{}^{\nu})} \,\bar\Gamma
&= \gamma^{(6)}
- \frac{1}{2}\G^{\mu\nu\lambda} u_{\mu}(\Ht\cdot u)_{\nu\lambda}\\
&\qquad\qquad - \ove{16\sqrt{-g}} {\e^{\mu_1\cdots\mu_6}}
		(\Ht\cdot u)_{\mu_1\mu_2}(\Ht\cdot u)_{\mu_3\mu_4} \G_{\mu_5\mu_6}.
\end{split}
\ee
So that $(1+\bar\Gamma)/2$ is the projector of rank 16,
and that
\be
\bar\Gamma^2= 1\,,\qquad \tr{\bar\Gamma}=0,
\ee
where
\be\label{gamma6}
\Gamma_\mu=E_\mu{}^A\Gamma_A\,,\qquad \gamma^{(6)}=\frac 1{6!\sqrt{-g}}\epsilon^{\mu_1\cdots\mu_6}\Gamma_{\mu_1\cdots\mu_6}\,.
\ee

\subsection{M5--brane action in the dual formulation}\label{subsecPSTdualM5}

In this paper, a covariantised M5-brane action in the dual formulation
is constructed with the help of $5$ auxiliary scalar fields $a^s(x).$
The index $s,$ as well as other from at the end of lower-case Roman alphabets,
labels the different auxiliary scalar fields, and
is chosen to be $s=0,1,2,3,4.$
This is just a choice of numbering and should
not be confused with spacetime indices.
The projector matrices associated to
the auxiliary fields are given by\footnote{The projector matrix $\PI$
was called $\P$ in \cite{Pasti:2009xc} and \cite{Ko:2013dka}.
The choice made in this paper is purely
because of
typesetting.
For each projector matrix
there is an induced projector, to be defined later, which are written using a calligraphic style.
We simply have no access to calligraphic version of $\P.$}
\be\label{projector5a}
P^\n_\m = \pa_\m a^r Y^{-1}_{rs}\pa^\n a^s,\qquad
\PI_\m^\n = \d_\m^\n - P_\m^\n,\qquad
P_\m^\n\pa_\n a^s = \pa_\m a^s,
\ee
where $Y^{-1}_{rs}$ is the matrix inverse of
\be
Y^{rs} = \pa_\m a^r \pa_\n a^s g^{\m\n}.
\ee
The projector $P$ has rank $5$
whereas the projector $\PI$ has rank $1.$

It is also convenient to define a vector
\be
\begin{split}
\l^\m
&=-\ove{5!}\ove{\sqrt{-g}}\e_{s_0 s_1 s_2 s_3 s_4} \zeta_{\m_0}^{s_0}\zeta_{\m_1}^{s_1}\zeta_{\m_2}^{s_2}\zeta_{\m_3}^{s_3}\zeta_{\m_4}^{s_4}\e^{\m\m_0\m_1\m_2\m_3\m_4},
\end{split}
\ee
where
\be
\e_{s_0 s_1 s_2 s_3 s_4}
=
\begin{cases}
1 & \textrm{even permutation of 01234}\\
-1 & \textrm{odd permutation of 01234}\\
0 & \textrm{otherwise}
\end{cases}
\ee
and $\zeta_\m^s \equiv \pa_\m a^s.$
It is related to the projector by the following identity
\be\label{lamb5aprojiden}
\PI_\m^\n = \frac{g_{\m\r}\l^\r\l^\n}{g_{\s\h}\l^\s\l^\h}
\equiv \frac{g_{\m\r}\l^\r\l^\n}{(\l)^2},
\ee
where
we have denoted $(\l)^2\equiv g_{\s\h}\l^\s\l^\h,$
which is not to be confused with the $\m = 2$ component of $\l^\m.$
The proof
of this identity and other discussions
related to the projectors will be discussed later
after we present some tools for calculations.

The M5-brane action in the dual formulation  
with PST covariantisation
in the 11d supergravity background constructed as a main result
of this paper
is given by
\begin{align}\label{covdual15action}
S_{cov-dual-M5}
=& \int_{\mathcal{M}_6} d^6x \lrsbrk{-\sqrt{-g}\sqrt{\det\lrbrk{\d_\m^\n + (H\cdot v)_\m{}^\n } }  + \frac{\sqrt{-g}}{4} (\Ht\cdot v)^{\m\n}(H\cdot v)_{\m\n} } \nn\\
&
 +\hlf \int_{\mathcal{M}_6} \lrbrk{ C_6 + H_3 \wedge C_3 },
\end{align}
with 
\be
(\Ht\cdot v)_{\m\n} \equiv \Ht_{\m\n\r}v^\r,\qquad
(H\cdot v)_{\mu\nu} \equiv H_{\mu\nu\rho} v^\r,\qquad
v^\m = \frac{\l^\m}{\sqrt{(\l)^2}}.
\ee
For
definiteness, we have put the action \eq{covdual15action}
in the form which accommodates $(\l)^2>0.$
In fact, the action \eq{covdual15action} can
be brought to the form which allows both $(\l)^2>0$ and $(\l)^2<0.$
This can be made possible because
$v^\m$ appear in pair in each expression and hence
after expressing them in terms of $\l^\m,$
the square roots in the denominators always appear in pair
$\sqrt{(\l)^2}\ \sqrt{(\l)^2} = (\l)^2.$

Similar to the case of the original M5-brane action,
the M5-brane action in the dual formulation also has symmetries of type PST1 and PST2
in addition to the conventional abelian gauge symmetry for $B_2.$
In this case, The PST1 symmetry is given by
\be\label{PST1dualM5}
\d a^s = 0,\qquad
\d B_{\m\n} = \pa_{[\m} a^r\pa_{\n]} a^s\psi_{rs}(a^w),
\ee
where $\psi_{rs}(a^w)$ are functions of auxiliary fields $a^s.$
Although semi-local, this symmetry allows the equation of motion
to be reduced to the nonlinear self-duality condition
\be\label{nonlinEOM}
(\Ht\cdot v)_{\m\n} = \cV_{\m\n}
,
\ee
where
\be
\cV^{\m\n}
\equiv 2\frac{\d\sqrt{\det(\d_\r^\s + (H\cdot v)_\r{}^\s)}}{\d (H\cdot v)_{\m\n}}.
\ee
The semi-locality of the PST1 symmetry
is analogous to its counterpart seen
in a PST covariantised version \cite{Maznytsia:1998xw,Maznytsia:1998yr} 
of chiral boson theory in two-dimensions \cite{Floreanini:1987as}.
The PST2 symmetry is given by
\be
\label{PST2dualM5}
\d a^s = \vphi^s,\qquad
\d B_{\m\n} = \hlf v_\r\vphi^r Y^{-1}_{rs}\pa_\s a^s \frac{\e^{\m'\n' \r\s\l\t}}{\sqrt{-g}}(\cV_{\l\t}-(\Ht\cdot v)_{\l\t})g_{\m\m'}g_{\n\n'},
\ee
where $\vphi^s(x)$ are arbitrary functions.
This symmetry ensures that the fields $a^s(x)$ are arbitrary.
By
following the analysis of \cite{Bandos:2014bva},
the dynamical system of the action \eq{covdual15action} is separated into two branches:
that with $(\l)^2>0,$ and that with $(\l)^2<0.$
These two branches are disconnected because there is no
non-singular PST2 transformation which can move the system
from one branch to the other
without passing through the forbidden region $(\l)^2 = 0,$
in which the action becomes singular.

In order for the second order field equation of $B$ to be gauge equivalent
to non-linear self-duality equation \eq{nonlinEOM},
the semi-local PST1 symmetry has to be a gauge symmetry.
This is the case when the Noether's charge vanishes \cite{Henneaux:1992ig,Bekaert:1998yp}.
Noether's current of the PST1 symmetry \eq{PST1dualM5} is given by
\be
j^\r
= \frac{1}{2}\psi_{rs}\pa_\m a^r \pa_\n a^s (\cV^{\m\n}-(\Ht\cdot v)^{\m\n})v^{\r},
\ee
which
is conserved on-shell. The
form of the Noether's current makes it clear that Noether's charge vanishes when $\l^0 = 0.$
In general, the analysis in each branch has to be done separately \cite{Bandos:2014bva}.
In the $(\l)^2 > 0$ branch, one can always use PST2 symmetry to gauge-fix $a^s = x^s$
giving $\l^0 = 0,$ which in turn implies that PST1 symmetry is a gauge symmetry
and can be used to ensure that the second order field equation of $B$
is equivalent to the non-linear self-duality condition.
On the other hand, throughout the $(\l)^2<0$ branch, 
the Noether's charge does not vanish.
So in this branch, the PST1 symmetry is a global symmetry,
and hence the non-linear self-duality condition
is not obtainable from gauge-fixing 
the second order field equation.

The M5-brane action in the dual formulation 
is also invariant under the kappa symmetry \eqref{PSTkappa},
which instead of $\d_\k a = 0$
we have $\d_\k a^s = 0.$
Additionally, $\bar\G$ for this theory is given via
\be
\begin{split}
\lrbrk{\sqrt{\det\lrbrk{\d_\m^\n + (H\cdot v)_\m{}^\n}}}\bar\G
&\\
&\!\!\!\!\!\!\!\!\!\!\!\!\!\!\!\!\!\!\!\!\!\!\!\!\!\!\!\!\!\!\!\!\!\!\!\!\!\!\!\!\!\!\!\!\!\!\!\!\!\!\!\!\!\!\!\!\!\!\!\!\!\!\!\!\!\!\!\!
=\g^\6 + \hlf v_\m(H\cdot v)_{\n\r}\g^\6\G^{\m\n\r} + \ove{16\sqrt{-g}}\e^{\m_1\cdots\m_6}(H\cdot v)_{\m_1\m_2}(H\cdot v)_{\m_3\m_4}
\G_{\m_5\m_6},
\end{split}
\ee
which also satisfies
\be
\bar\G^2 = 1,\qquad
\tr\bar\G = 0.
\ee

\section{Derivations}\label{sec:derive}
In this section, we present the derivation of the M5-brane action
in dual formulation \eq{covdual15action}.
By using differential form language,
the construction and the study of the properties
of the action is naturally made possible.
Therefore, let us first develop the necessary tools
before working on the construction.

\subsection{Mathematical preliminary: induced linear transformation}

The M5-brane action
in dual formulation presented by eq.\eq{covdual15action}
requires $5$ auxiliary scalar fields $a^s,\ s=0,1,2,3,4,$
which appear in the action only via their gradients $\z^s\equiv da^s.$
In principle, the study of the action \eq{covdual15action}
can be done by directly making use of five 1-forms $\z^s.$
However, we find it more convenient to study by using
projectors $P^\m_\n, \PI^\m_\n$ incorporated 
into differential form language.
This can be done by using the idea of induced linear transformation.
Let us now give a quick review on this idea.
See for example \cite{Renteln:2013sdz, Winitzki}
for more information.
The discussions and examples presented in the following
can be easily generalised and made suitable
for the context and purpose of this paper.
Readers who are familiar with this mathematical language
may wish to read this subsection quickly to find out the convention we used.

Let $V$ be a vector space with $V^*$ its dual space.
Consider a linear map
\be
T:V\to V.
\ee
The transpose of $T$ is given by a linear map
\be
T^\dag: V^*\to V^*
\ee
such that
\be
\z(T(v)) = (T^\dag\z)(v),\qquad\forall v\in V,\forall\z\in V^*.
\ee

Given two or more linear maps $V\to V,$
a multilinear map on products of $V$ can be introduced.
For example, consider two linear maps $T:V\to V,$ and $S:V\to V.$
An induced transformation $\bigw^2 T\bigw S$
is a multilinear map
\be
\begin{split}
\bigw^2 T\bigw S:\otimes^3 V&\to\bigw^3 V\\
(v_1, v_2, v_3)&\mapsto Tv_1\w T v_2\w S v_3.
\end{split}
\ee
Other induced maps, for example $\bigw^3 T,\ \bigw T\bigw S\bigw T,$ etc. can also be defined in a similar manner.
A ``trace'' is given by the sum of all possible permutations of the induced transformations.
For example,
\be
\tr(\bigw^2 T\bigw S) = \bigw T\bigw T\bigw S + \bigw T\bigw S \bigw T+ \bigw S\bigw T\bigw T.
\ee
These maps are totally antisymmetric.
For example
\be
\begin{split}
\tr(\bigw^2 T\bigw S)(v_1, v_2, v_3)
&= -\tr(\bigw^2 T\bigw S)(v_1, v_3, v_2)
=\tr(\bigw^2 T\bigw S)(v_3, v_1, v_2)\\
=-\tr(\bigw^2 T\bigw S)(v_3, v_2, v_1)
&=\tr(\bigw^2 T\bigw S)(v_2, v_3, v_1)
=-\tr(\bigw^2 T\bigw S)(v_2, v_1, v_3).
\end{split}
\ee
The ``trace'' satisfies a binomial expansion property
\be
\bigw^n(T+S) = \sum_{r=0}^n \tr(\bigw^r T\bigw^{n-r}S).
\ee
It is clear that the constructions on a dual vector space
can be defined in a similar way.

Let us now consider a useful identity.
For example, let $T:V\to V,$ and $S:V\to V$
be a linear map, and let $F$ be a 3-form.
Then, it can be shown that
\be
\tr(\bigw^2 T^\dag\bigw S^\dag)(F) = F\circ\tr(\bigw^2 T\bigw S),
\ee
where $\circ$ is the symbol for function composition.
To avoid future clutter of notation,
we will simply drop the symbols $\dag$ and $\circ$
as it should be clear from the context where these symbols should appear.
So we may simply write the above equation as
\be
\tr(\bigw^2 T\bigw S)(F) = F(\tr(\bigw^2 T\bigw S)).
\ee

In the subsequent subsections, we start by reviewing
the construction of a covariantised quadratic
action for chiral $2-$form in dual formulation \cite{Maznytsia:1998xw,Maznytsia:1998yr},
in which the covariantisation is made possible with
the help of an auxiliary $4-$form.
We then show the potential issues which
could possibly 
prevent the extension of the action
to a complete M5-brane action in the dual formulation.
Our next goal is not to thoroughly investigate whether these issues
are truly problematic, let alone to try to resolve them. We simply limit the study to a
special case of auxiliary fields which avoid these potential issues. This choice will make
it evident that the extension to a complete M5-brane action in the dual formulation is
possible.

\subsection{Quadratic dual action of a six-dimensional chiral 2-form 
theory with an auxiliary 4-form}\label{sec:4formform}
Let us give a review and analysis of the quadratic action
of a chiral 2-form in six dimensions
in a dual formulation with an auxiliary 4-form
constructed by \cite{Maznytsia:1998xw,Maznytsia:1998yr}. We translate the presentation into
differential form language.
We use the convention that exterior derivatives
and interior products act from the right,
and that a $p-$form is expressed as
\be
A_p = \ove{p!}dx^{\m_1}\w\cdots\w dx^{\m_p}A_{\m_p\cdots\m_1},
\ee
and a Hodge star is given by
\be
*dx^{\m_1}\w\cdots dx^{\m_p}
=\frac{(-1)^{p+1}}{(6-p)!\sqrt{-g}}dx^{\m_{p+1}}\w\cdots\w dx^{\m_6}\e^{\n_{p+1}\cdots\n_{6}\m_1\cdots\m_p}g_{\m_{p+1}\n_{p+1}}\cdots g_{\m_{6}\n_{6}},
\ee
where $x^\m,\ \m=0,1,\cdots,5$ are 6d coordinates,
and $g$ is the determinant of the 6d metric.
Let us denote the field strength of a chiral 2-form as
\be
F = dB,
\ee
and define
\be
\cF = F - *F.
\ee

The action constructed by \cite{Maznytsia:1998xw,Maznytsia:1998yr} made use of an auxiliary 4-form $\chi_4$
which appears in the
action via the Hodge dual of its field strength:
\be
\tilde{\l} = *d\chi.
\ee
This naturally gives rise to the projectors
\be
\Pt = \frac{g^{-1}(\tilde{\l})\otimes\tilde{\l}}{g^{-1}(\tilde{\l},\tilde{\l})},\qquad
\PIt = \mathbbm{1}-\Pt, 
\ee
where $\mathbbm{1}$ is the identity map.
Here the inverse metric $g^{-1}$
takes the role of a linear map which maps a one-form to a vector.
The induced linear transformations are defined as
\be  \label{cPtcPIt}
\cPt \equiv \tr(\bigw\Pt\bigw^2\PIt),\qquad
\cPIt \equiv \tr(\bigw^3\PIt),\qquad
\cI \equiv \tr(\bigw^3\mathbbm{1}).
\ee
They satisfy the following identities
\be
\cPt+\cPIt = \cI,\qquad
\cPt\bigw\cI = \cI\bigw\cPIt,\qquad
\cI\bigw\cPt = \cPIt\bigw\cI.
\ee
It can be shown that for any 3-form $A_3,$
\be
\cPIt A_3= -\ove{\tilde{\l}^2}i_{g^{-1}\tilde{\l}}(\tilde{\l}\w A_3).
\ee

With the above setup, we can write the 6d chiral 2-form action of
\cite{Maznytsia:1998xw,Maznytsia:1998yr} as
\be\label{dual15w4form}
S = \int \ove{2}F\w\cPIt\cF.
\ee
The variations with respect to the $2-$form field
and auxiliary $4-$form field are given by
\be
\d_{(B)}S = \int \d B\w d\cPIt\cF -\hlf \int d(\d B\w(2\cPIt\cF-F)),
\ee
\be
\begin{split}
\d_{(\chi)}S
&= \int\frac{1}{2\tilde{\l}^2}\d \tilde{\l}\w\cPt\cF\w i_{g^{-1}\tilde{\l}}\cPt\cF\\
&= \hlf \int \d\chi\w d*\lrbrk{\ove{\tilde{\l}^2}\cPt\cF\w i_{g^{-1}\tilde{\l}}\cPt\cF}
-\int d\lrbrk{\ove{2\tilde\l^2}\d\chi\w*\lrbrk{\cPt\cF\w i_{g^{-1}\tilde{\l}}\cPt\cF}}.
\end{split}
\ee
So the field equations for $B$ and $\chi$ are
\be\label{EOMB-4form}
d\cPIt\cF = 0,
\ee
\be\label{EOMchi-4form}
d*\lrbrk{\ove{\tilde{\l}^2}\cPt\cF\w i_{g^{-1}\tilde{\l}}\cPt\cF} = 0.
\ee
Apart from the tensor gauge symmetry of $B,$
the action \eq{dual15w4form} also has tensor gauge symmetry
for $\chi,$ as well as PST1 and PST2 symmetries.
The tensor gauge variation for $\chi$ is given as an
exterior derivative of a $3-$form gauge parameter,
which is reducible. Out of the $20$ components of
$3-$form gauge parameter, only $(20-(15-(6-1))) = 10$
are independent.
The PST1 symmetry of the action \eq{dual15w4form}
is given by
\be\label{PST1-4form}
\d B = \ove{\sqrt{\tilde{\l}^2}}i_{g^{-1}\tilde{\l}}\Psi,\qquad
\d\chi = 0,
\ee
where the parameter $\Psi$ satisfies
\be
\cL_{g^{-1}\tilde{\l}}\lrbrk{\ove{\sqrt{\tilde{\l}^2}}i_{g^{-1}\tilde{\l}}\Psi} = 0,
\ee
where $\cL_{g^{-1}\tilde{\l}}$ is the Lie derivative
along the vector field $g^{-1}\tilde{\l}.$
The PST2 symmetry is given by
\be\label{PST2dual15w4form}
\d B = i_\xi \cPIt\cF,\qquad
\d\chi = i_\xi *\tilde{\l},
\ee
where the parameter $\xi$ is an arbitrary vector field.
The variation $\d\chi$ implies the variation on $\tilde{\l}$
as
\be\label{PST2lambt}
\d \tilde{\l} = g(\cL_\xi g^{-1}\tilde{\l}) + \textrm{div }\xi\ \tilde{\l}.
\ee
Here, the metric $g$ takes a role of a linear map, which maps a vector to a one-form.

In general, PST1 symmetry is used in order to reduce
the second order field equation \eq{EOMB-4form}
to self-duality equation. In order to do so,
PST1 has to be a gauge symmetry.
However, PST1 symmetry \eq{PST1-4form} is semi-local (see for example \cite{Bekaert:1998yp,Bandos:2014bva} for similar issues),
which means that it can either be a gauge symmetry
or a global symmetry.
In order for the PST1 symmetry to be a gauge symmetry,
its Noether's charge has to vanish (see for example \cite{Henneaux:1992ig}).
The Noether's charge is given by the 5d spatial integral of $j^0,$
where
\be
j = -*\lrbrk{\ove{\sqrt{\tilde\l^2}}i_{g^{-1}\tilde\l}\Psi\w\cPIt\cF}.
\ee
In order for $j^0$ to vanish, one demands that
\be\label{j0-4form}
dt\w i_{g^{-1}\tilde\l}\Psi\w\cPIt\cF = 0.
\ee
By adopting the viewpoint similar to that of \cite{Bandos:2014bva},
one may expect that the dynamical system is separated into two branches:
that with $g^{-1}(\tilde\l,\tilde\l) > 0,$
and that with $g^{-1}(\tilde\l,\tilde\l) < 0.$
The task is to determine the branch in which
the condition \eq{j0-4form} is satisfied.
Let us now give an analysis on this.

Consider the transformation
\be\label{PST2plusTensor}
\d\chi = i_\xi d\chi + di_\xi\chi.
\ee
The first term on the RHS is a PST2 transformation,
while the second term is a tensor gauge transformation
whose parameter is identified with $i_\xi\chi.$
The transformation \eq{PST2plusTensor}
is simply given by a Lie derivative acting on $\chi.$
Therefore, it is well-known that an associated finite
transformation is given by
\be
\chi^{(h)} = \ove{4!}d(x^\m+h\xi^\m)\w d(x^\n+h\xi^\n)\w d(x^\r+h\xi^\r)\w d(x^\s+h\xi^\s)\chi^{(0)}_{\s\r\n\m}(x+h\xi),
\ee
where $h$ is a parameter along the integral curve of $\xi.$
One then obtains
\be
\begin{split}
\tilde\l^{(h)}
&=\ove{4!}dx^\r\frac{\e^{\n_5\n_1\n_2\n_3\n_4\m_5}}{\sqrt{-g}}g_{\n_5\r}\pa_{\n_1}(x^{\m_1}+h\xi^{\m_1})
\pa_{\n_2}(x^{\m_2}+h\xi^{\m_2})\times\\
&\qquad\qquad\pa_{\n_3}(x^{\m_3}+h\xi^{\m_3})\pa_{\n_4}(x^{\m_4}+h\xi^{\m_4})
\pa_{\m_5}\chi_{\m_4\m_3\m_2\m_1}(x+h\xi).
\end{split}
\ee
Given $\chi^{(0)},$ and $\xi,$ it can be seen that $g^{-1}(\tilde\l^{(h)},\tilde\l^{(h)})$
varies smoothly in $h.$
Using this result and the fact that the dynamical system is not defined at $g^{-1}(\tilde\l,\tilde\l) = 0,$
one concludes that the dynamical system is separated into two branches:
that with $g^{-1}(\tilde\l,\tilde\l) > 0,$
and that with $g^{-1}(\tilde\l,\tilde\l) < 0.$
It is not possible to connect these two branches without passing through
the region with $g^{-1}(\tilde\l,\tilde\l) = 0.$

In the $g^{-1}(\tilde\l,\tilde\l) > 0$
branch, one can use the combined transformation \eq{PST2plusTensor}
to gauge fix $\chi$ to, say
\be\label{choicechifix}
\chi = -x^0 dx^{1234},
\ee
which gives $j^0 = 0,$ and hence PST1 is a gauge symmetry
making the field equations \eq{EOMB-4form}-\eq{EOMchi-4form} to be gauge equivalent
to self-duality condition $\cF = 0.$
On the other hand, in the $g^{-1}(\tilde\l,\tilde\l) < 0$ branch,
one always have $j^0\neq 0.$ Therefore,
one does not obtain self-duality condition in this branch.

By counting the number of components, one may expect that
the action \eq{dual15w4form} has a potential issue with
PST2 symmetry \eq{PST2dual15w4form}.
If one makes use of reducible tensor gauge symmetry of $\chi,$
i.e. by gauge-fixing,
then the number of remaining independent components of $\chi$ is $15-10 = 5.$
So 5 out of 6 independent components of PST2 parameter $\xi$ are used
to completely gauge away the remaining components of $\chi.$
The remaining 1 independent PST2 parameter
could potentially remove 1 degree of freedom of $B.$
The predicted removal of component of $B$ by gauge-fixing PST2 symmetry
is not desired and could be considered as an issue.

In order to make sure, one will need to give an explicit
analysis to see whether the issue actually arises.
However, we do not intend to pursue this investigation through the end.
Let us simply give a remark that
in an example of gauge-fixing to a
non-manifest covariant theory,
the issue does not seem to arise.
Suppose that one has used the combined PST2 and tensor gauge transformation
to gauge-fix $\chi$ to
\be\label{gaugex5-4form}
\chi = \ove{5!}\e_{\una\unb\unc\und\une 5}x^\una dx^{\unb}\w dx^{\unc}\w dx^{\und}\w dx^{\une},
\ee
where underlined lower case Roman indices $\una,\unb,\ldots$
take the values $0,1,2,3,4.$
Next, by demanding that the combined
diffeomorphism, PST2, and tensor gauge transformation
do not change this gauge, one obtains
\be
\begin{split}
0
&=dx^{\una}\w dx^{\unb}\w dx^{\unc}\w dx^{\und}\lrbrk{\ove{4!}\e_{\una\unb\unc\und\une 5}(\xi^\une + \e^{\une}) + \ove{5}\ove{3!}\pa_{\und}(\e^\unf\e_{\une\una\unb\unc\unf 5}x^{\une})+\ove{3!}\pa_{\und}\g_{\unc\unb\una}}\\
&\qquad+dx^{\una}\w dx^{\unb}\w dx^{\unc}\w dx^{5}\lrbrk{\ove{5}\ove{3!} \pa_5(\e^\und\e_{\une\una\unb\unc\und 5}x^{\une})
+\frac{4}{3!}\pa_{[5}\g_{\unc\unb\una]}},
\end{split}
\ee
where $\g_{\m\n\r}$ is the parameter for the tensor gauge transformation of $\chi,$
and $\e$ is the parameter for the diffeomorphism transformation.
This condition is solved by
\be\label{fix4form-soln}
\g_{5\unb\una} = 0,\qquad
\g_{\unc\unb\una} + \ove{5}\e^{\und}\e_{\une\una\unb\unc\und 5}x^\une = 0,\qquad
\xi^\una = -\e^\una,
\ee
which is a special solution.
Note that there is no condition which specifies $\xi^5$
component of the PST2 transformation.
Naively, this component could potentially kill a degree of freedom of
$B_2.$ However, an explicit analysis shows
that this is not the case.
Under the combined diffeomorphism and PST2 transformation,
and after imposing \eq{fix4form-soln},
one obtains
\be\label{B-moddiff-4formform}
\begin{split}
\d B
&= \cL_\e B - \e^\una i_{\pa_\una}\cPIt\cF\\
&= \cL_\e B - \frac{2}{g_{55}}\e^\unp dx^{\unm\unn} g_{5[5}\cF_{\unp\unn\unm]},
\end{split}
\ee
which is clear that $\xi^5$ does not enter
and hence no degree of freedom is unintentionally removed.

The fact that the extra component of PST2 parameter
does not appear in the above example is interesting.
However, we leave it as a future work to investigate in a more general setup
whether the extra component of PST2 parameter
would remain unharmful.

\subsection{4-form to 5 scalars}
The paper \cite{Maznytsia:1998xw,Maznytsia:1998yr} derives the
quadratic dual action of a 6d chiral 2-form with an auxiliary 4-form
by starting from the covariant quadratic action
of a 6d chiral 2-form  
with an auxiliary scalar $a(x),$
and then applying a dualisation technique on the auxiliary scalar.
The process gives rise to a quadratic dual action
of a 6d chiral 2-form 
such that an auxiliary field must appear through
a $1-$form $\tilde{\l}$ satisfying the condition $d*\tilde{\l} = 0.$
The converse of the Poincare's lemma
then gives $\tilde{\l} = *d\chi,$
for an arbitrary $4-$form $\chi.$
This is how the auxiliary 4-form appears in the paper \cite{Maznytsia:1998xw,Maznytsia:1998yr}.

As discussed in the previous subsection,
it is still unclear whether there is an issue when using an auxiliary $4-$form.
So we only follow the above procedure up to a certain step,
and then put in some restrictions.
In particular, we follow the procedure
up to the step
in which the condition $d*\tilde{\l} = 0$ is obtained.
Imposing some restrictions then means
that
a suitable decomposition has to be made
on the solution of $\tilde{\l}.$
For example, one might wish to use a Helmholtz decomposition
and then restricting to a special case by turning off some fields
in the decomposition.
However, for the problem at hand, Helmholtz decomposition is not suitable.
To find a more suitable decomposition,
we make use of a geometrical interpretation.
Recall that in a PST covariantised theory,
an auxiliary scalar field $a$ appears in the action
via a $1-$form $\z = da.$
The geometrical interpretation is that
$\z$ describes a normal to 5D hypersurfaces $a=const.$
For the dual theory, however, the condition $\tilde{\l} = db$
for some scalar field $b$ cannot be imposed as it contradicts to
$d*\tilde{\l} = 0.$ So a different interpretation has to be made.
An alternative description of 5D hypersurfaces
is given by wedge product of five $1-$forms.
Therefore, the decomposition we look for is to
decompose $*\tilde{\l}$ into a wedge product of five $1-$forms
plus some other terms. It turns out that this problem is related to a decomposability problem
in the context of exterior algebra.

So let us first discuss this problem in exterior algebra.
Consider a 6-dimensional dual vector space $V^*.$
We would like to investigate the conditions
in which a 5-form $*\l$ can be written as
a wedge product of $5$ 1-forms.
It turns out that this is always possible.
For the proof, let us closely follow the arguments in the reference
\cite{Hitchin2003}, adopted to the case at hand.
Let us define a linear map
\be
\begin{split}
T:V^*&\to\w^6 V^*\\
w&\mapsto (*\l)\w w.
\end{split}
\ee
Note that $\textrm{dim}(\textrm{im }T)\leq\textrm{dim}(\w^6 V^*) = 1.$
So from rank-nullity theorem, we have
$\textrm{dim}(\textrm{ker }T) \geq 5,$
which means that the kernel should consist of at least $5$ linearly independent 1-forms.
Let $w^0, w^1, w^2, w^3, w^4$ be linearly independent 1-forms
in the kernel. Then extend the set of these 1-forms
to a basis $w^0, w^1, w^2, w^3, w^4, w^5$ in $V^*.$
This allows us to write the 5-form $*\l$ as
\be
\begin{split}
*\l &= \l_{01234}w^0\w w^1\w w^2\w w^3\w w^4 + \l_{01235}w^0\w w^1\w w^2\w w^3\w w^5\\
&\qquad+\l_{01245}w^0\w w^1\w w^2\w w^4\w w^5 + \l_{01345}w^0\w w^1\w w^3\w w^4\w w^5\\
&\qquad+\l_{02345}w^0\w w^2\w w^3\w w^4\w w^5 + \l_{12345}w^1\w w^2\w w^3\w w^4\w w^5.
\end{split}
\ee
Since $w^0, w^1, w^2, w^3, w^4\in \textrm{ker }T,$
we have
\be
(*\l)\w w^0 = (*\l)\w w^1 = (*\l)\w w^2 = (*\l)\w w^3 = (*\l)\w w^4 = 0.
\ee
So
\be
\l_{12345} = \l_{02456} = \l_{01345} = \l_{01245} = \l_{01235} = 0.
\ee
This leaves us with
\be
*\l = \l_{01234}w^0\w w^1\w w^2\w w^3\w w^4,
\ee
which indeed shows that any 5-form in $\bigw^5 V^*$
where $\textrm{dim} V^* = 6$ is always
decomposable in terms of a wedge product
of 5 1-forms.

The above proof works for tensors but not necessarily
tensor fields as in the case of our concern.
Nevertheless, we suppose
that after some suitable restrictions, if any,
the above result can also be applied.
This means that the theorem suggests that
a generic 5-form $*\l$
can be written as, modulo some possible restrictions when generalising from tensors to tensor fields,
\be\label{gen5formdecQ}
\begin{split}
*\l
&= l(x) w^0(x)\w w^1(x)\w w^2(x)\w w^3(x)\w w^4(x)\\
&= l(x) \ove{5!}\e_{s_0 s_1 s_2 s_3 s_4}w^{s_0}(x)\w w^{s_1}(x)\w w^{s_2}(x)\w w^{s_3}(x)\w w^{s_4}(x).
\end{split}
\ee
Next, applying the condition $d*\l = 0$ gives
\be\label{dldweqn}
\begin{split}
0 &= \ove{5!}\e_{s_0 s_1 s_2 s_3 s_4}dl(x)\w w^{s_0}(x)\w w^{s_1}(x)\w w^{s_2}(x)\w w^{s_3}(x)\w w^{s_4}(x)\\
&\qquad+l(x) \ove{4!}\e_{s_0 s_1 s_2 s_3 s_4}w^{s_0}(x)\w w^{s_1}(x)\w w^{s_2}(x)\w w^{s_3}(x)\w d w^{s_4}(x),
\end{split}
\ee
which is implied by
\be\label{dldwsoln}
dl = d w^s = 0,
\qquad s\in\{0,\cdots,4\}.
\ee
Note
that this is not necessarily a general solution.
Our goal is simply to look for a possible reduction
of a 4-form, use it as auxiliary field in the covariantisation,
and see if it solves the issues discussed in subsection \ref{sec:4formform}.
So a special solution to eq.\eq{dldweqn} is sufficient for our purpose.
However, it will be interesting for future investigation
to see what a general solution looks like, and whether it would also
eventually serve the purpose.
The solution \eq{dldwsoln} is solved by
\be
l = const. \equiv -1,\qquad w^s = da^s,\qquad s\in\{0,\cdots,4\}.
\ee
So
\be\label{slamb5a}
*\l = -da^0\w da^1\w da^2\w da^3\w da^4
\ee
We will make use of this decomposition in the construction
of covariant formulation of dual M5-brane.
This means that the solution $\tilde{\l}=*d\chi$ to $d*\tilde{\l} = 0$
is restricted as
\be\label{eq4formto5scalars}
\tilde{\l}|\equiv \tilde{\l}\bigg|_{\chi = -da^0\w da^1\w da^2\w da^3 a^4} = -*\lrbrk{da^0\w da^1\w da^2\w da^3\w da^4} = \l.
\ee

\subsection{Quadratic dual action of a 6d chiral 2-form with five auxiliary scalars}
In this subsection, we construct and show in detail that
the restriction made by eq.\eq{eq4formto5scalars}
allows the successful covariantisation of the quadratic action of the 6d chiral 2-form in the dual formulation.

Applying the restriction \eq{eq4formto5scalars}
to the action \eq{dual15w4form}
gives
\be\label{dual15w4formrest}
S = \int \hlf F\w\cPIt|\cF,
\ee
where
\be
\cPt| \equiv \tr(\bigw\cPt|\bigw^2\PIt|),\qquad
\cPIt| \equiv \tr(\bigw^3\PIt|),
\ee
with
\be
\Pt| = \frac{g^{-1}(\tilde{\l}|)\otimes\tilde{\l}|}{g^{-1}(\tilde{\l}|,\tilde{\l}|)}
\equiv \frac{g^{-1}(\l)\otimes\l}{g^{-1}(\l,\l)}
,\qquad
\PIt| = \mathbbm{1}-\Pt|,
\ee
and the symbol
$|$ denotes the restriction \eqref{eq4formto5scalars} to the choices of five auxiliary scalars.
Since $\l$ is expressed by eq.\eq{slamb5a},
the action \eq{dual15w4formrest} requires $5$ auxiliary scalar fields
$a^s, s=0,1,2,3,4$ via the gradients $da^s.$

The projector matrices $P_\m^\n, \PI_\m^\n$ defined in the eq.\eq{projector5a}
can also be written as
\be  \label{Pzeta}
P = Y^{-1}_{rs} g^{-1}(\z^r)\otimes \z^s
\equiv g^{-1}(\z_s)\otimes\z^s,\qquad
\PI = \mathbbm{1} - P,
\ee
where $\z^s = da^s,\ \z_r\equiv Y^{-1}_{rs}\z^s.$
The fact that ranks of $P$ and $\PI$
are $5$ and $1,$ respectively, are symbolically represented by
\be
\bigw^6 P = 0 = \bigw^2\PI.
\ee
Then
\be
\begin{split}
\bigw^6\mathbbm{1}
&= \bigw^6(P+\PI)\\
&= \tr(\bigw^5 P\bigw\PI).
\end{split}
\ee
Next, let us denote
\be  \label{cPcPI}
\cP\equiv \bigw^3 P,\qquad
\cPI\equiv \tr(\bigw^2 P\bigw\PI),\qquad
\cI\equiv\bigw^3\mathbbm{1}.
\ee
They satisfy the following identities
\be\label{cPwcI}
\cP+\cPI = \cI,\qquad
\cP\bigw\cI = \cI\bigw\cPI,\qquad
\cI\bigw\cP = \cPI\bigw\cI.
\ee

Let us now verify the identity \eq{lamb5aprojiden}.
By direct calculation, this gives
\be
g^{-1}(\l)\otimes \l = -\det Y (\mathbbm{1} - P)
\ee
So
\be
g^{-1}(\l,\l) = -\det Y (6 - 5) = -\det Y,
\ee
and hence
\be
\begin{split}
\Pt| 
&= \frac{g^{-1}(\l)\otimes\l}{g^{-1}(\l,\l)}\\
&= \mathbbm{1} - P\\
&= \PI,
\end{split}
\ee
as required.
Using identity \eq{lamb5aprojiden}, with the definitions \eqref{cPtcPIt} and \eqref{cPcPI}, the action \eq{dual15w4formrest} can be rewritten as
\be\label{dual15w5a}
S = \hlf\int F \w \cP\cF.
\ee

Next, let us discuss the computation of the variation
of the action \eq{dual15w5a}.
The variation of the action with respect to $B$
can be computed using the identities \eq{cPwcI}
as well as
\be
*\cP = \cPI*.
\ee
The variation is done as follows
\be\label{dSdAresult}
\begin{split}
\d_{(B)} S
&= \int \hlf \d F \w\cP\cF + \hlf F\w\cP\d F - \hlf F\w\cP*\d F\\
&= \int \hlf \d F\w(\cP\cF - \cPI F - \cP*F)\\
&= \int \d F\w\cP \cF - \hlf\d F\w F\\
&= \int \d B \w d(\cP\cF) - \hlf d(\d B\w(2\cP\cF-F)).
\end{split}
\ee
As for the variation of the action with respect to $a^s,$
it is useful to
first consider the variation of the projector $P = g^{-1}(\z_s)\otimes\z^s$ with respect to $a^s:$
\be\label{dPda}
\d_{(a)} P = \PI g^{-1}(\d \z^s)\otimes \z_s + g^{-1}(\z_s)\otimes\PI\d \z^s.
\ee
Then from an identity
\be
\cP \cF = -2 \cF + \z^s\w i_{g^{-1}\z_s}\cF,
\ee
we can use eq.\eq{dPda} to read off
\be
\begin{split}
\d_{(a)}\cP\cF
&= \z_s\w i_{\PI g^{-1}(\d \z^s)}\cF + \PI\d \z^s\w i_{g^{-1}(\z_s)}\cF.
\end{split}
\ee
Further calculation gives
\be
\d_{(a)}\cP\cF
=(1+*)(\PI\d \z^s\w i_{g^{-1}\z_s}\cF).
\ee
Then the variation of the action with respect to $a^s$
can be done as follows
\be\label{dSdaresult}
\begin{split}
\d_{(a)}S
&= \hlf\int F\w(1+*)(\PI\d \z^s\w i_{g^{-1}\z_s}\cF)\\
&= -\hlf\int \d \z^s\w\cP\cF\w i_{g^{-1}\z_s}\cP\cF\\
&= -\int \d a^s i_{g^{-1}\z_s}(\cP\cF)\w d(\cP\cF)
+ \hlf\int d(\d a^s\cP\cF\w i_{g^{-1}\z_s}\cP\cF).
\end{split}
\ee
In the last step, we use the identity
\be\label{diLFiden}
d(\cP F_1\w i_{g^{-1}\z_s}\cP F_2)
= d\cP F_1\w i_{g^{-1}\z_s}\cP F_2 + i_{g^{-1}\z_s}\cP F_1\w d\cP F_2,
\ee
which is valid for any $3-$forms (as well as $3-$form superfields) $F_1$ and $F_2.$
This can be shown by using 
Leibniz rules for $d$ and $i_{g^{-1}\z_s},$
Cartan's magic formula,
and the identity
\be
\cL_{g^{-1}\z_s}P = [g^{-1}\z_s,g^{-1}\z_r]\otimes \z^r,
\ee
where $[\cdot,\cdot]$ is a Lie bracket.

Combining eq.\eq{dSdAresult}, and eq.\eq{dSdaresult} gives
\be\label{dSdAdSda}
\begin{split}
\d_{(B)} S+\d_{(a)}S
&= \int (\d B - \d a^s i_{g^{-1}\z_s}(\cP\cF)) \w d(\cP\cF)\\
&\qquad- \int \hlf d(\d B\w(2\cP\cF-F))
+ \hlf\int d(\d a^s\cP\cF\w i_{g^{-1}\z_s}\cP\cF).
\end{split}
\ee
This gives the field equations for $B,$ and $a^s:$
\be\label{EOMAsol}
d(\cP\cF) = 0,
\ee
\be\label{EOMauxsol}
i_{g^{-1}\z_s}(\cP\cF) \w d(\cP\cF) = 0,
\ee
So clearly, the field equations for $a^s$
are implied by the field equations for $B.$
The variation \eq{dSdAdSda} can also be used to
read off the PST1 and PST2 symmetries.
PST1 symmetry is only due to the transformation of $B.$
So it should satisfy
\be\label{gtcond}
\cPI d\d B = 0,\qquad \d a^{s} = 0.
\ee
The form of $\d B$ which solves this condition is given by
\be
\begin{split}
\d B
&= \tr(\bigw^2 P)\F\\
&= \hlf\psi_{rs}da^{s}\w da^r,
\end{split}
\ee
where $\psi_{rs} = i_{g^{-1}\z_s}i_{g^{-1}\z_r}\F.$
Then
\be
\cPI d\d B = \hlf\PI d\psi_{rs}\w da^{s}\w da^{r}.
\ee
So the condition \eq{gtcond} implies that
\be
\PI d\psi_{rs} = 0.
\ee
Note that since
\be
\PI da^r = 0,
\ee
$\psi_{rs}$ should be a function of $a^w.$
So PST1 symmetry is given by
\be
\d B = \hlf\psi_{rs}(a^w)da^{s}\w da^{r},\qquad
\d a^s = 0,
\ee
which is semi-local.
The analogous form \cite{Cherkis:1997bx,Maznytsia:1998xw,Maznytsia:1998yr}
can be seen in the covariant version
of Floreanini-Jackiw $d=2$ chiral boson theory \cite{Floreanini:1987as}.
As for PST2 symmetry, it involves the variations of $B,$ and $a^s.$
This symmetry can easily be read off from the equation \eq{dSdAdSda}
giving
\be
\d a^s = \vphi^s,\qquad
\d B = \vphi^s i_{g^{-1}\z_s}(\cP\cF).
\ee
The PST2 symmetry is used to ensure that the auxiliary fields $a^s$
are arbitrary.
Therefore, it is not surprising that the field equations of
the auxiliary scalars, eq.\eq{EOMauxsol}
are implied by the field equations of $B,$ eq.\eq{EOMAsol}.

To
complete the analysis of the action \eq{dual15w5a},
we need to investigate the possible case in which
the second order field equation \eq{EOMAsol}
is equivalent to self-duality condition.
For this let us closely
follow the analysis given by \cite{Bandos:2014bva}.
We first note that the action and field equations
are singular when $g^{-1}(\l,\l) = 0.$ This separates
the dynamical system into two branches:
that with $g^{-1}(\l,\l)>0,$ and that with $g^{-1}(\l,\l)<0.$
To see that the two branches are really separated,
one considers a generic integral curve generated by PST2 transformation.
Let $h$ be a parameter along the integral curve,
then given $a^s = a^s_{(0)}$ at $h=0,$ the scalars evolve as
\be
a^s_{(h)} = a^s_{(0)} + h\vphi^s.
\ee
Then
\be
g^{-1}(\l_{(h)},\l_{(h)}) = g\det Y_{(h)},
\ee
where $\l_{(h)}$ is given as in eq.\eq{slamb5a}
with $a^s$ replaced by $a^s_{(h)},$
and $\det Y_{(h)}$ is the determinant of a matrix
\be
Y^{rs}_{(h)} = (\pa_\m a_{(0)}^r + h\pa_\m\vphi^r)g^{\m\n}(\pa_\n a_{(0)}^s + h\pa_\n\vphi^s).
\ee
It can then be seen that
along the curve, the value of $g^{-1}(\l,\l)$ varies smoothly.
Therefore, if a curve connects a point with $g^{-1}(\l,\l)>0,$
and another point with $g^{-1}(\l,\l)<0,$
then it should inevitably pass through
the singular region with $g^{-1}(\l,\l) = 0.$
The two branches of the dynamical system
will need to be studied separately
to see which branch would give self-duality condition.
For this, one needs the PST1 symmetry to be a gauge symmetry.
A criteria for this is that
the PST1 symmetry is a gauge symmetry
if its Noether's charge vanishes \cite{Henneaux:1992ig}.
The Noether's current is
\be
j = -*((\tr(\w^2 P)\F)\w\cP\cF).
\ee
It can be shown that $j^0 = 0$
when $\l^0 = 0.$
This is the case only in the $g^{-1}(\l,\l) > 0$ branch,
in which PST2 gauge transformation can be used
to gauge-fix $a^s = x^s,$ giving $\l^0 = 0.$
So in this branch PST1 is a gauge symmetry,
and can be used to gauge fix
the equation \eq{EOMAsol} to a self-duality condition
\be
F = *F.
\ee
On the other hand, PST1 is a global symmetry
in the $g^{-1}(\l,\l) < 0$ branch.
Therefore, in this branch
one does not obtain self-duality condition
from a gauge-fixing of field equation.

By fixing the gauge
\be\label{gaugex5-5scalars}
a^s = x^s,
\ee
and demanding that the combined PST2 and 6d diffeomorphism transformation
does not modify this gauge condition,
one obtains
\be
\vphi^s = -\e^s,
\ee
where $\e^\m$ is a 6d diffeomorphism parameter.
Under the combined PST2 and 6d diffeomorphism transformation,
$B_2$ transforms as
\be\label{B-moddiff-5scalarsform}
\begin{split}
B
&= \cL_\e B - 2\frac{\e^s}{g_{55}}dx^{pq} g_{5[5}\cF_{sqp]},
\end{split}
\ee
which is a modified diffeomorphism transformation rule
of the non-manifest covariant chiral 2-form with quadratic action
in dual formulation.
It also exactly agrees with eq.\eq{B-moddiff-4formform}
which was intended to come from exactly the same theory.

Having
reviewed quadratic action of chiral $2-$form in dual formulation
covariantised using an auxiliary $4-$form \cite{Maznytsia:1998xw,Maznytsia:1998yr}
in subsection \ref{sec:4formform},
and having presented the alternative covariantisation
of the theory using 5 auxiliary scalars in this subsection,
let us give a remark on whether these two
actions are related.
A 4-form field has $15$ components.
However, the reducible tensor gauge symmetry
reduces the number of components to be $15 - (20-(15-(6-1))) = 5$
which agrees with the number of scalar fields we introduced.
In fact, an example of the relationship can be seen from eq.\eq{eq4formto5scalars}
which suggests that $5$ auxiliary scalars
give a particular choice of $\chi,$
i.e. $\chi = -da^0\w da^1\w da^2\w da^3 a^4.$
Given $a^s,$ other equivalent choices of $\chi$ can also be made
for example
\be\label{choice2chi}
\chi = -a^0 da^1\w da^2\w da^3\w da^4,
\ee
which is related to the choice of eq.\eq{eq4formto5scalars} by the tensor gauge transformation
$\d\chi = d(a^0a^4 da^1\w da^2\w da^3).$
Furthermore, in the gauge \eq{gaugex5-5scalars} for $a^s$,
the choice \eq{choice2chi} reduces to eq.\eq{choicechifix}
which is the corresponding gauge choice of $\chi.$
Another notable choice of $\chi$ in terms of given $a^s$
is
\be\label{gaugex5-4form-before}
\chi = \ove{5!}\e_{\una\unb\unc\und\une 5}x^{\una}dx^{\unb}\w dx^{\unc}\w dx^{\und}\w dx^{\une},
\ee
which is related to the choice
\eq{eq4formto5scalars} by the tensor gauge transformation
\be
\d\chi = \frac{4}{5}d(a^4 a^{[0} da^1\w da^2\w da^{3]}).
\ee
In the gauge \eq{gaugex5-5scalars},
the choice \eq{gaugex5-4form-before} reduces to eq.\eq{gaugex5-4form}.

Having seen explicit examples of the relationship between the two types of auxiliary fields,
a natural question to ask is
whether the 5 auxiliary fields
give a parametrisation of the independent components
of the auxiliary 4-form field.
In order for the parametrisation to valid,
one needs to check at the level of the field equation
to see if the equations \eq{EOMB-4form}-\eq{EOMchi-4form}
would reduce, after setting $\chi$ for example
as in eq.\eq{eq4formto5scalars},
to the equations \eq{EOMAsol}-\eq{EOMauxsol}.
While we still do not have a direct check,
there is a supporting evidence that this could be the case;
the full check will be left as a future work.
Previously, we have investigated that for $\chi$ in the gauge
\eq{gaugex5-4form}, where the theory reduces to
a non-manifest covariant chiral 2-form with quadratic action
in dual formulation, the equations \eq{EOMB-4form}-\eq{EOMchi-4form}
are equivalent to self-duality condition $F = *F.$
For $5$ auxiliary scalar in the gauge \eq{gaugex5-5scalars},
which corresponds to $\chi$ in the gauge \eq{gaugex5-4form},
the equations \eq{EOMAsol}-\eq{EOMauxsol} also reduce to
self-duality condition $F = *F.$
Furthermore, the modified diffeomorphism transformations \eq{B-moddiff-4formform}
for the theory with gauge-fixed $4-$form agrees with
the one \eq{B-moddiff-5scalarsform}
for the theory with gauge-fixed $5$ scalars.

\subsection{Nonlinear dual action of a 6d chiral 2-form with five auxiliary scalars}\label{derivenonlinaction}
Having obtained a quadratic action for a 6d chiral 2-form theory with five auxiliary scalars
and shown 
that it indeed has desirable properties,
let us now extend it to a non-linear action. By looking, for example, at
the non-manifestly covariant M5-brane action in the dual formulation \cite{Ko:2016cpw},
it is natural to write down the action
\be
S = \int d^6 x\sqrt{-g}\lrbrk{-\sqrt{\det(\d_\m^\n + (F\cdot v)_\m{}^\n)} + \ove{4}(\tilde{F}\cdot v)^{\m\n}(F\cdot v)_{\m\n}},
\ee
where
\be
(F\cdot v)_{\m\n} = F_{\m\n\r}v^\r,\qquad
(\Ft\cdot v)_{\m\n} = \Ft_{\m\n\r} v^\r,
\ee
\be
v_\m = \frac{\l_\m}{\sqrt{g^{-1}(\l,\l)}},\qquad
\Ft^{\m\n\r} = \ove{6\sqrt{-g}}\e^{\m\n\r\l\s\t}F_{\l\s\t}.
\ee

Let us show that this action indeed has desirable properties.
We first start from the variation of the action,
\be\label{varySdualnonlin}
\begin{split}
\d_{(B)}S + \d_{(a)}S
=&-\int\lrbrk{\d B + \d a^s i_{g^{-1}\z_s}*\cPI(W-*F)}\w d*\cPI(W-*F)
\\
&+\hlf \int d\lrbrk{\d B\w(2*\cPI(W-*F)+F)}\\
&+\hlf \int d\lrbrk{\d a^s*\cPI(W-*F)\w i_{g^{-1}\z_s}*\cPI(W-*F)},
\end{split}
\ee
where
\be
W = \ove{3!}dx^\m\w dx^\n\w dx^\r W_{\r\n\m},
\ee
\be
W_{\m\n\r} = \frac{\lrbrk{1+\hlf (F\cdot v)_{\l\s}(F\cdot v)^{\l\s}}F_{\m\n\r}+\frac{3}{2}(F\cdot v)_{[\m|\s}(F\cdot v)^{\s\l}F_{\l|\n\r]}}{\sqrt{\det(\d_\m^\n + (F\cdot v)_\m{}^\n)}}.
\ee
The variation \eq{varySdualnonlin}
can be obtained by using tools and steps
similar to the quadratic action, in particular
the identity \eq{diLFiden}.
We also use the identity
\be\label{WWFFiden}
i_{g^{-1}\z_s}*\cPI W\w*\cPI W
= i_{g^{-1}\z_s}*\cPI F\w*\cPI F.
\ee

Using the variation of the action, the field equations for $B,$ and $a^s$
are
\be\label{EOMnonlinB}
d*\cPI(W-*F) = 0,
\ee
\be\label{EOMnonlina}
i_{g^{-1}\z_s}*\cPI(W-*F)\w d*\cPI(W-*F)
= 0,
\ee
which is clear that the field equations for $a^s$ are implied by the field equations for $B.$
Next, the PST1 symmetry reads
\be
\d B = \hlf\psi_{rs}(a^w)da^{s}\w da^{r},\qquad
\d a^s = 0.
\ee
In the case where PST1 is a gauge symmetry,
it will be used to gauge fix the field equation for $B$
to give
\be
\cPI(W-*F) = 0,
\ee
or
\be\label{nlsds}
i_{g^{-1}v}*F = i_{g^{-1}v}W.
\ee 
Next, PST2 Symmetry is given by
\be
\d a^s = \vphi^s,\qquad
\d B = - \vphi^s i_{g^{-1}\z_s}*\cPI(W-*F),
\ee
which can be used to ensure that the $a^s$
are indeed auxiliary.

By
a similar analysis to the previous subsection,
it can also be concluded that in the $g^{-1}(\l,\l)>0$ branch
the second order field equation \eq{EOMnonlinB} can be gauge-fixed to give
nonlinear self-duality condition \eq{nlsds},
whereas in the $g^{-1}(\l,\l)<0$ branch,
the second order field equation
is not gauge equivalent to nonlinear self-duality condition.

Having shown that the action \eq{varySdualnonlin}
has desirable properties, it is then natural to
extend this action to a complete M5-brane action in the dual formulation
\eq{covdual15action}. As for the symmetries,
it can be easily checked that
the conventional abelian gauge symmetry 
and the PST1 symmetry are not modified, whereas
the PST2 symmetry is modified by having all
$F$ promoted to $H = F+C.$
As well as the bosonic symmetries,
the couple of the M5-brane to an 11d supergravity background
also enjoys a local
fermionic symmetry called kappa-symmetry.
The check of kappa-symmetry can easily be done by following
the standard techniques used for example in \cite{Aganagic:1997zq, Bandos:1997ui}.

\section{Gauge-fixing auxiliary fields}\label{sec:gaugefixtononcov}
In \cite{Ko:2016cpw}, the non-manifest
covariant M5-brane action
in the dual formulation
coupled to 11d supergravity background
was presented and shown that the theory is justified.
The checks were done by using constrained analysis,
comparison of on-shell actions,
and double dimensional
reduction to D4-brane.

It can be shown that the 
covariant M5-brane action
in the dual formulation
coupled to 11d supergravity background
presented in section
\ref{subsecPSTdualM5}
can be reduced to the action of \cite{Ko:2016cpw}. 
Let us start by using the 
PST2 symmetry \eq{PST2dualM5}
of the action \eq{covdual15action}
to fix the gauge
\be\label{gaugefixa}
a^s = x^s,\qquad
\textrm{so}\qquad
\pa_\m a^s = \d_\m^s.
\ee
This gives
\be
\l^\m = \frac{\d^\m_5}{\sqrt{-g}},\qquad
\l_\m = \frac{g_{\m 5}}{\sqrt{-g}},
\ee
and hence
\be
v^\m = \frac{\d^\m_5}{\sqrt{g_{55}}},\qquad
v_\m = \frac{g_{\m 5}}{\sqrt{g_{55}}},
\ee
\be
(H\cdot v)_{\m\n} = \frac{H_{\m\n 5}}{\sqrt{g_{55}}},
\qquad
(\Ht\cdot v)^{\m\n} = \ove{3!}\frac{\e^{\m\n\r\s\l\t}}{\sqrt{-g}}H_{\s\l\t}\frac{g_{\r 5}}{\sqrt{g_{55}}}.
\ee
It can easily be seen that in the gauge \eq{gaugefixa}
the action \eq{covdual15action}
reduces to the
non-manifestly covariant
M5-brane action in the dual formulation
constructed in \cite{Ko:2016cpw}.
Under the combined local transformation of PST2
and 6d diffeomorphism $\d x^\m = \xi^\m(x)$,
the auxiliary fields transform as
\be
\d a^s(x) = \xi^\m(x)\pa_\m a^s(s) + \vphi^s(x) = \xi^s(x) + \vphi^s(x).
\ee
This combined transformation should not 
modify the gauge-fixing condition \eq{gaugefixa}.
So the PST2 gauge parameter should be chosen to be
\be
\vphi^{s}(x) = -\xi^s(x),
\ee
in which case, the combined local transformation
on $B_{\m\n}$
is given by
\be\label{moddiffeo}
\d B_{\m\n} =
\xi^\r\pa_\r B_{\m\n} + 2\pa_{[\n} \xi^\r B_{\m]\r} 
- \xi^q\lrbrk{4\ove{g_{55}}g_{5[5}H_{\m\n q]}
+\e_{\m\n q mn5}\lrbrk{-\hlf \frac{\cV^{mn}}{\sqrt{g_{55}}}\sqrt{-g}}},
\ee
We see that this is simply a modified diffeomorphism
transformation obtained
from the analysis in \cite{Ko:2016cpw}.
In particular, the modification only appears in the $\xi^m$ directions
of the components
$\d B_{mn}.$

In \cite{Ko:2016cpw}, after presenting and analysing the non-manifestly covariant
M5-brane action in the dual formulation,
the analyses of comparison of on-shell actions, and of double dimensional
reduction to D4-brane were discussed.
These analyses do not require the use of auxiliary fields.
So one can safely say that these analyses are indeed also valid for
the covariant M5-brane action in the dual formulation.

So far in the literature, there are three alternative descriptions
of the complete M5-brane actions: (i) the original M5-brane action \cite{Aganagic:1997zq, Bandos:1997ui}.
(ii) the M5-brane action in the 3+3 formulation \cite{Pasti:2009xc,Ko:2013dka},
and (iii) the M5-brane action in the dual formulation constructed in \cite{Ko:2016cpw} and this paper.
Although the off-shell actions from different descriptions are different from one another,
it was shown that they all agree on-shell.
The consequence of this is for example that
these actions give the same value for the
the tension of a string soliton solution.

The double dimensional reduction of M5-brane action in the
dual formulation is done by compactifying
one direction on M5-brane on a circle.
The theory directly reduces to a D4-brane theory
coupled to a ten-dimensional type IIA
supergravity background \cite{Douglas:1995bn}, \cite{Green:1996bh}
without the need to make any further dualisation.

\section{Conclusion}\label{sec:conclude}
In this paper we have presented the
covariant M5-brane action in dual formulation
coupled to 11d supergravity background.
The covariantisation of this theory
is made possible by using $5$ auxiliary fields.
It can be shown that by gauge-fixing PST2
symmetry, the constructed action can be reduced
to the non-manifestly covariant version
constructed and analysed in \cite{Ko:2016cpw}.
It is then evident that the action constructed in this paper
inherits   
some properties from the the one constructed in \cite{Ko:2016cpw}. 

We
have demonstrated that at the quadratic level of the action,
the covariantised action with $5$ auxiliary scalar fields, eq.\eq{dual15w4formrest},
can be obtained by
replacing auxiliary 4-form field in the action of \cite{Maznytsia:1998xw,Maznytsia:1998yr},
which is written using differential form as eq.\eq{dual15w4form},
by using eq.\eq{eq4formto5scalars}.
Although the number of independent components
of the auxiliary 4-form field is the same as that of the $5$
auxiliary scalar fields,
the substitution using
eq.\eq{eq4formto5scalars} at the action level
does not necessarily mean that the $5$ auxiliary scalar fields
are result from a parametrisation of the auxiliary 4-form field.
In fact, this has to be studied at the level of field equations.
By gauge-fixing to non-manifestly covariant theory,
we found a supporting evidence that this might be the case.
However, it is still not enough to conclude in favour or against
this. We leave the full verification as a future work.

In \cite{Ko:2013dka}, 
the covariant M5-brane action
coupled to 11d supergravity background
is constructed with the help of 3 auxiliary fields.
This result and the result of our paper
suggests that PST covariantisation
using more than 1 auxiliary scalar field
is also possible.
However, the paper \cite{Ko:2015zsy}
attempted
to obtain a covariant M5-brane action
using 2 auxiliary fields,
but did not succeed.

As for self-dual fields in other dimensions,
it is also interesting to investigate whether
covariantisation using more than 1 auxiliary scalar field
is possible.
For example, we expect that
for a chiral 4-form theory in 10 dimensions,
the version with 9 auxiliary scalar fields
is possible and is dual to the usual PST version
with 1 auxiliary field.
At the moment, this is just only an anticipation.
We plan to work on this as a future work to see
if this is really the case.

The PST covariantisation
has also been used in duality-symmetric theories \cite{Zwanziger:1970hk,Deser:1976iy,Schwarz:1993vs},
which are the theories generalising
the duality transformation between electric fields
and magnetic fields.
More recently,
as a way to investigate and study counterterms
in supergravity and string theory effective action,
the non-linearisation of duality-symmetric action
in 4d is constructed and analysed in \cite{Bossard:2011ij, Carrasco:2011jv}.
The theory is covariantised in \cite{Pasti:2012wv}.
The covariantisation of the
dual theory of this using 3 auxiliary scalar fields
will be reported separately.

\subsection*{Acknowledgements}
We are very grateful to Dmitri Sorokin for various helpful remarks and comments on the
manuscript.

The work of P.V. is supported by
the Office of the Higher Education Commission (OHEC)
and the Thailand Research Fund (TRF)
through grant number MRG6080136.
P.V. would also like to thank his mentor, Burin Gumjudpai,
for helps and advices especially on the
management of this grant. Sh.-L. K. would like to thank his friends in the journey, especially members of Institute for Fundamental Study and people around Naresuan university. 


\providecommand{\href}[2]{#2}\begingroup\raggedright\endgroup

\end{document}